\documentclass[aps,prl,prb,twocolumn,showpacs,superscriptaddress,groupedaddress]{revtex4-1}
\usepackage{graphicx}
\usepackage{physics}
\usepackage{amssymb}
\usepackage{amsmath}
\usepackage{latexsym}
\usepackage{ulem}
\usepackage{color}
\usepackage{dcolumn}
\usepackage{subfigure}
\usepackage[T1]{fontenc}
\usepackage[utf8]{inputenc}
\usepackage[english,french]{babel}
\usepackage[colorlinks=true,linkcolor=blue,citecolor=blue]{hyperref} %
\usepackage{bm}        
\usepackage{amssymb}   
\usepackage{booktabs}

\usepackage[capitalize]{cleveref}

\usepackage{lipsum}
\usepackage[dvipsnames]{xcolor}

\begin{document}

\graphicspath{{figures/}}

\definecolor{airforceblue}{rgb}{0.36, 0.54, 0.66}

\title{Crossing over from flat band superconductivity to conventional superconductivity}

\author{M. Thumin}%
 \email{maxime.thumin@neel.cnrs.fr}
 \author{G. Bouzerar}%
 \email{georges.bouzerar@neel.cnrs.fr}
\affiliation{Université Grenoble Alpes, CNRS, Institut NEEL, F-38042 Grenoble, France}%
\date{\today}

\begin{abstract}
Over the past ten years, flat band (FB) or geometric superconductivity has become a major issue in condensed matter physics due to the significant technological benefits it could offer. Observations of this unconventional form of superconductivity are unfortunately still very limited, and significant efforts are being made to search for candidate materials.
Most existing theoretical studies focus on systems with strictly non-dispersive bands, which, from an experimental point of view, represents an extremely difficult technological constraint to achieve. It is therefore crucial to understand to what extent this constraint can be relaxed. In other words, to what extent can superconductivity in flat bands survive weak perturbations?
The main objective of the present study is precisely to answer this essential question in detail. 

\end{abstract} 

\maketitle


\section{Introduction}

The standard BCS theory of superconductivity suggests that the higher the density of states at the Fermi energy $\rho_F$, the higher the critical temperature according to $T_c \propto t \, e^{-1/\rho_F|U|}$ \cite{BCS}. A natural first choice for achieving higher critical temperatures is to rely on van Hove singularities (VHS).
In the half-filled square lattice where the Fermi energy coincide with that of the VHS, one finds $T_c \propto t \, e^{-\sqrt{t/|U|}}$ \cite{Dupuis}.
Another interesting idea emerged in the 1990s \cite{Gap_linear_1990,Gap_linear_1994} and became a key stone in the last decade \cite{Batrouni_CuO2,Aoki2020}: flat bands (FB). The superconductivity of dispersionless bands (diverging density of states) revealed that the pairing $\Delta$ could be proportional to the attractive interaction strength $|U|$ \cite{Gap_linear_1990,Gap_linear_1994,BCS_FB,Batrouni_sawtooth} as well as the critical temperature $T_c \propto |U|$ \cite{Volovik_T_linear, Peotta_Nature}.  
Later, the nature of this unconventional form of superconductivity was elucidated when it was shown that the superfluid weight $D_s$ is proportional to $|U|$ and to the quantum metric \cite{Peotta_Nature, Peotta_Lieb, Iskin}, a key quantity in the framework of quantum geometry \cite{Provost_metric, Berry_5_years}. This lead to the proposal that real-space decimation could be an effective strategy to strengthen the superconducting phase \cite{Thumin_boost_QM}. 
The peculiar scalings of $\Delta$, $D_s$ and more importantly $T_c$ allow us to get rid of the exponentially suppressed scaling predicted by the BCS theory, opening a new path to the room-temperature superconductivity.
\\
Until recently, flat bands were realized mostly in artificial systems such as in Lieb or Kagome optical lattices \cite{Kagome_optical_lattice,Lieb_optical_lattice, Lieb_optical_lattice2,Sawtooth_optical_lattice} or in the field of photonics \cite{photonic_Lieb_Kagome_Honeycomb,photonic_metallic_Kagome,photonic_metallic_Lieb}, however the experimental realization in real compounds is still lacking. Today, two-dimensional heterostructures stand out as a major platform suitable for realizing FBs in transition metal dichalcogenides bilayers such as TaS$_2$ \cite{TaS2_exp,TaS2_th}, WSe$_2$ \cite{FB_WSe2}, MoSe$_2$ \cite{MoS2_WS2_WSe2}, in heterolayers \cite{heterobilayers}, or even in semiconducting heterostructures such as In$_{0. 53}$Ga$_{0.47}$As/InP \cite{III-V_SC}. 
Since 2018, FB superconductivity has had an experimental realization in twisted bilayer graphene \cite{Cao2018_1,Cao2018_2} at magic angle $\theta=1.05^\circ$, and even more recently in WSe$_2$ at $\theta=5.0^\circ$ \cite{FB_supra_WSe2}.
However, there is still a grey area that needs to be clarified. In fact, it is crucial to know \textit{how flat the band needs to be in order to observe this unconventional form of superconductivity}. This question is essential from the point of view of experimental realization. Equivalently, \textit{would the introduction of a mechanism that makes the band weakly dispersive have a dramatic effect on FB superconductivity?} Such a study would also help to understand how the crossover between standard BCS behavior (intraband superconductivity) and the FB superconductivity (interband superconductivity) takes place. This is the main goal of the present study.
We emphasize that this study is part of a very topical issue, since recently the effect on tuning the bandwidth of a quasi-FB (QFB) to probe the proximity effect was studied on the twisted bilayer graphene \cite{proximity_effect_FB}.
\begin{figure}[h!]
\centering
\includegraphics[scale=0.4]{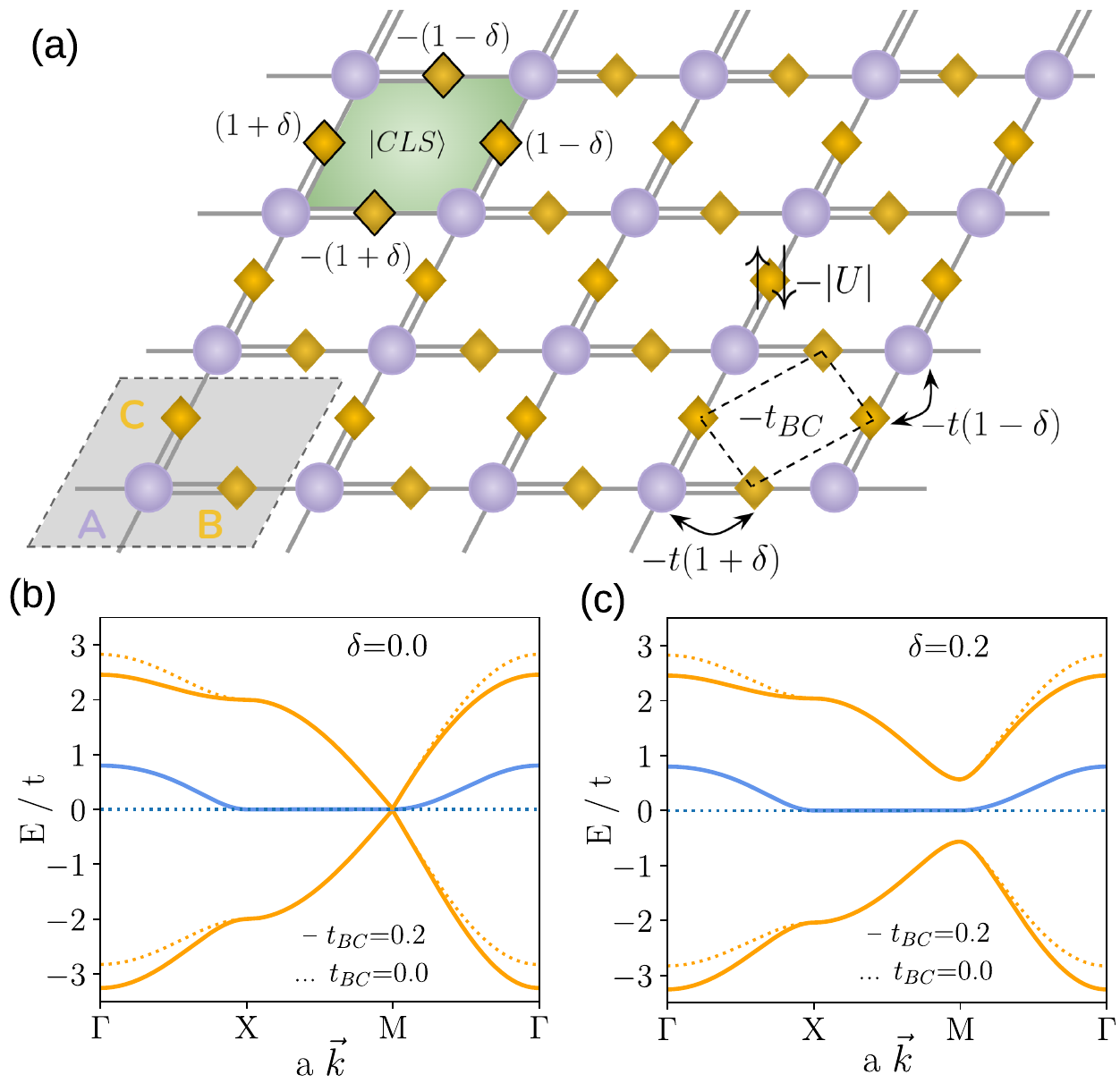} 
\caption{
(a) Sketch of the dimerized Lieb lattice and a compact localized state (CLS), only the non-zero weights of the CLS are shown (black contoured $B$ and $C$-atoms). 
(b) (resp. (c)) Dispersions for $|U| = 0$ along the $\Gamma-X-M-\Gamma$ path in the Brillouin zone for $\delta=0$ (resp. $\delta=0.2$). In each panel, $t_{BC}$ is 0 (dotted lines) or $0.2$ (continuous lines).
}
\label{Fig. 1}
\end{figure}
\\
To address superconductivity in weakly dispersive bands, we consider the Lieb lattice, a two-dimensional decorated square lattice with three orbitals per unit cell ($A,B$ and $C$) as displayed in Fig.~ \ref{Fig. 1}(a). The Lieb lattice corresponds to CuO$_2$ planes in cuprates, and is known to host a FB at $E=0$. 
In the half-filled system, the presence of repulsive electronic interaction leads to a ferrimagnetic ground state \cite{Lieb_Lattice_Origine} and to geometric FB superconductivity when the interaction are attractive \cite{Peotta_Lieb}.
The existence of the FB in this lattice relies on the biparticity and thus can be broken by considering intra-sublattice hoppings. A gap between the FB and the dispersive bands can be tuned by introducing a dimerization of the lattice (distorted lattice).
This system becomes therefore a versatile platform to answer the questions raised above. 
Remark that in the present study we consider only the case of trivial bands (Chern number is zero).
However, several studies have focused their attention on the interplay between superconductivity and topology leading to interesting bound inequalities for the superfluid weight \cite{Bound_inequality_roy,Peotta_Nature,Hofmann_QFB,Torma_band_geometry_berry}.

\section{Model and methods}
The electrons are described by the attractive dimerized Hubbard model whose Hamiltonian reads,
\begin{equation}
\begin{split}
    \hat{H} = \sum_{i\lambda,j\eta, \sigma} t^{\lambda\eta}_{ij} \; \hat{c}_{i\lambda, \sigma}^{\dagger} \hat{c}_{j \eta, \sigma} - \mu \hat{N} \underset{\hat{H}_{int}}{\underbrace{- |U| \sum_{i\lambda} \hat{n}_{i\lambda\uparrow}\hat{n}_{i\lambda\downarrow}}} 
    .
    \label{H_exact}
\end{split}
\end{equation}
$\hat{c}_{i \lambda \sigma}^{\dagger}$ creates an electron in the $\lambda = A,B,C$ orbital of the $i$-th cell with spin $\sigma=\uparrow,\downarrow$. 
The only non-zero hoppings correspond to those between nearest neighbors $-t(1\pm\delta)$, $\delta$ being the dimerization amplitude, and between next nearest neighbors $t_{BC}$ as illustrated in Fig.~\ref{Fig. 1}(a).
$\mu$ is the chemical potential, $\hat{N}=\sum_{i\lambda} \hat{c}_{i \lambda \sigma}^{\dagger}\hat{c}_{i \lambda \sigma}$ the particle number operator and $|U|$ the interaction strength. All energy are expressed in unit of $t$.
\\
Here, the superconductivity is addressed within the multi-band Bogoliubov-de-Gennes (BdG) approximation, which has proven extremely reliable in FB and QFB systems for the calculation of the occupations, the pairings, the one-body correlation functions as well as the superfluid weight in one and two-dimensional systems \cite{Peotta_Lieb, Batrouni_CuO2,Batrouni_sawtooth, Batrouni_Designer_Flat_Bands,Hofmann_PRB,Chi_QMC,Peri_PRL,Arbeitman_PRL, Thumin_coherence_length}.
We emphasize that these numerous studies have demonstrated that the BdG approach is not only qualitatively reliable but also quantitatively reliable even in the case of one-dimensional systems for which quantum and thermal fluctuations have the most dramatic impacts.
\\
Within this approach,
the many body part of the Hamiltonian becomes,
\begin{equation}
\begin{split}
\hat{H}_{int} \overset{BdG}{\simeq} \sum_{i\lambda}
-&|U| \langle\hat{n}_{i\lambda,\downarrow}\rangle  \hat{n}_{i\lambda,\uparrow} -|U|\langle\hat{n}_{i\lambda,\uparrow}\rangle  \hat{n}_{i\lambda,\downarrow} \\
+&  \, \Delta_{i\lambda} \hat{c}^{\dagger}_{i\lambda,\uparrow}\hat{c}^{\dagger}_{i\lambda,\downarrow} 
+ \Delta_{i\lambda}^* \hat{c}_{i\lambda,\downarrow}\hat{c}_{i\lambda,\uparrow}  \\ 
+ & |U|  \langle\hat{n}_{i\lambda,\uparrow}\rangle  \langle \hat{n}_{i\lambda,\downarrow} \rangle  + \frac{|\Delta_{i\lambda}|^2}{U} \; , 
\end{split}
\end{equation}
$\langle\hat{n}_{i\lambda,\uparrow}\rangle$ and $\Delta_{i\lambda}=-|U|\langle \hat{c}_{i\lambda,\downarrow}\hat{c}_{i\lambda,\uparrow} \rangle$ are respectively the occupations and the pairings that are calculated self-consistently. In what follows, the cell index $i$ is dropped (translational invariance). We define the filling factor as $\nu=\sum_{\lambda\sigma}\langle\hat{n}_{\lambda,\sigma}\rangle$ and set it to $3$ (half-filling) unless otherwise stated.
Average values are calculated within the grand canonical ensemble, $\langle \hat{\mathcal{O}} \rangle = \text{Tr} [ \hat{\mathcal{O}} \, e^{-\hat{H}^{BdG}/k_BT} ] / \text{Tr} [ e^{-\hat{H}^{BdG}/k_BT} ]$, where $\hat{H}^{BdG}$ reads,
\begin{eqnarray}
    & \hat{H}^{BdG}(\textbf{q}) = \sum_\textbf{k}
    \hat{\Lambda}_\textbf{k}^\dagger 
    \,
    \hat{H}_\textbf{k}^{BdG}(\textbf{q})
    \,
    \hat{\Lambda}_\textbf{k}
    + C(\textbf{q})
    ,
    \\
    \nonumber 
    \\
    & \hat{H}_\textbf{k}^{BdG}(\textbf{q}) = 
    \begin{bmatrix}
         [\hat{h}_{\textbf{k-q}}^\uparrow]_{\lambda\eta}-V_\lambda \delta_{\lambda\eta}  &    \Delta_\lambda \delta_{\lambda\eta} \\ 
        \Delta^*_\lambda \delta_{\lambda\eta} &      -[\hat{h}_{\textbf{-k-q}}^{\downarrow}]^*_{\lambda\eta}+V_\lambda \delta_{\lambda\eta}  \\
    \end{bmatrix}
    .
\end{eqnarray}
Notice that the phase twist $\textbf{q}$ is introduced by Peierls substitution $\hat{c}_{i\lambda,\sigma} \longrightarrow \hat{c}_{i\lambda,\sigma} e^{i\textbf{qr}_{i\lambda}}$ to probe the superconducting phase. 
$\hat{\Lambda}_\textbf{k}^\dagger = \Big( \hat{\textbf{c}}_{\lambda \textbf{k-q} \uparrow}^{\dagger}, \hat{\textbf{c}}_{\lambda \textbf{-k-q} \downarrow} \Big)$ is the Nambu spinor, $h_\textbf{k}^\sigma$ is the Fourier transform of the tigh-binding part of Eq.~\eqref{H_exact}, and $V_\lambda = \mu+|U| \langle \hat{n}_{\lambda} \rangle/2$.
$C(\textbf{q})= \sum_{k \lambda} [\hat{h}^{\downarrow}_{\textbf{-k-q}}]_{\lambda\lambda} - V_\lambda(\textbf{q}) + \sum_{i \lambda} \frac{|\Delta_\lambda(\textbf{q})|^2}{|U|} + |U| \frac{\langle\hat{n}_{\lambda}(\textbf{q}) \rangle ^2}{4}$ a $\textbf{q}$-dependent quantity necessary for the calculation of the order parameter. Except for $C(\textbf{q})$, local quantities ($n_\lambda$, $\Delta_\lambda$) are calculated for $\textbf{q=0}$. 
The single particle Hamiltonian being time-reversal symmetric, we use $\hat{h}_{\textbf{-k-q}}^{\downarrow*}=\hat{h}_{\textbf{+k+q}}^{\uparrow}$
\\
The order parameter (superfluid weight) $D_s$ is defined as,
\begin{equation}
D_s^{\alpha\beta}(T)=\frac{1}{\mathcal{V}}\frac{d^2\Omega(\textbf{q})}{dq^{\alpha}dq^{\beta}} \Big|_{\textbf{q}=\textbf{0}} 
,
\label{def_Ds}
\end{equation}
where $\Omega(\textbf{q})=-k_BT \ln \text{Tr} [e^{-\hat{H}^{BdG}(\textbf{q})/k_BT}]$ is the grand canonical potential, and $\mathcal{V}$ is the volume of the system. Notice that the derivative is total, thus the 'constant' $C(\textbf{q})$ must be kept in the BdG decoupling \cite{Torma_revisiting, Batrouni_sawtooth} because of the implicit q-dependence of both the occupations and pairings.
Finally, because of lattice symmetry, $D_s^{xx}=D_s^{yy}$, hence we choose on $\alpha=\beta=x$ and note $D_s^{xx}=D_s$.

\section{Results and discussions}
\subsection{Spectrum and occupations (U=0)}
First, before we switch on the electron-electron interaction, we briefly discuss the spectrum and eigenstates of the non-interacting Hamiltonian. In the basis $\hat{\textbf{c}}_{\lambda k\sigma}^{\dagger}=(\hat{c}_{Ak\sigma}^{\dagger},\hat{c}_{Bk\sigma}^{\dagger},\hat{c} _{Ck\sigma}^{\dagger})$, it reads, 
\begin{equation}
    h_\textbf{k}^\sigma = 
    \begin{bmatrix}
        0   &  f_x^* & f_y^*  \\ 
        f_x &      0 & f_{xy} \\
        f_y & f_{xy} & 0      \\
    \end{bmatrix}
    ,
\end{equation}
where $f_\alpha = -2t\cos{(k_\alpha a /2)}-2ti\delta \sin{(k_\alpha a /2)}$, $\alpha=x,y$ and $f_{xy}=-4t_{BC}\cos{(k_x a /2)}\cos{(k_y a /2)}$.
The single-particle spectrum is depicted in Fig.~\ref{Fig. 1}(b,c) along the path $\Gamma-X-M-\Gamma$ in the Brillouin zone. Dispersions for both, $t_{BC}=0$ and $t_{BC}=0.2$ are shown for $\delta=0$ in panel (b) and $\delta=0.2$ in panel (c). 
At $t_{BC}=0$, the spectrum is well known $\{\varepsilon_\textbf{k}^{\pm} =\pm \sqrt{|f_x|^2+|f_y|^2}, \varepsilon_\textbf{k}^{FB}=0 \}$ 
and hosts a FB which is either gapless ($\delta=0$) or gapped ($\delta=0.2$) with a symmetric gap $2\sqrt{2}\delta\,t$. The relevant eigenstates of the flat band are Wannier functions also named compact localized states $|CLS\rangle$ whose weight is vanishing outside the plaquette they live on (see Fig.~\ref{Fig. 1}(a)).
The existence of the FB relies on the chiral symmetry (biparticity), which is broken when $t_{BC} \ne 0$. In this case, the FB turns into a quasi-flat band (QFB) with energy $\varepsilon_\textbf{k}^{QFB}=\frac{-2f_{xy}\Re{f_xf_y^*}}{(\varepsilon^+_\textbf{k})^2}$ (first order perturbation theory) and finite bandwidth $\mathcal{W}_{FB}=4t_{BC}$.
Note that for $t_{BC} \geqslant 0.5$ the definition of $\mathcal{W}_{FB}$ is ambiguous because of a band-touching which occurs at the $\Gamma$-point. 
\\
We remark that Lieb's uniform density theorem ($n_\lambda=1$) \cite{Th_Lieb_uniform} does not apply anymore, leading to a charge density wave at half-filling where the occupations are found larger on $B$ and $C$-orbitals (respectively $A$-orbitals) if 
$t_{BC} > 0$ (respectively $t_{BC} < 0$). More precisely, numerically, we find that $n_B = n_C \simeq 1 + 0.15 t_{BC}$ in the case of vanishing dimerization amplitude.
\\
\begin{figure}[ht]
\centering
\includegraphics[scale=0.5]{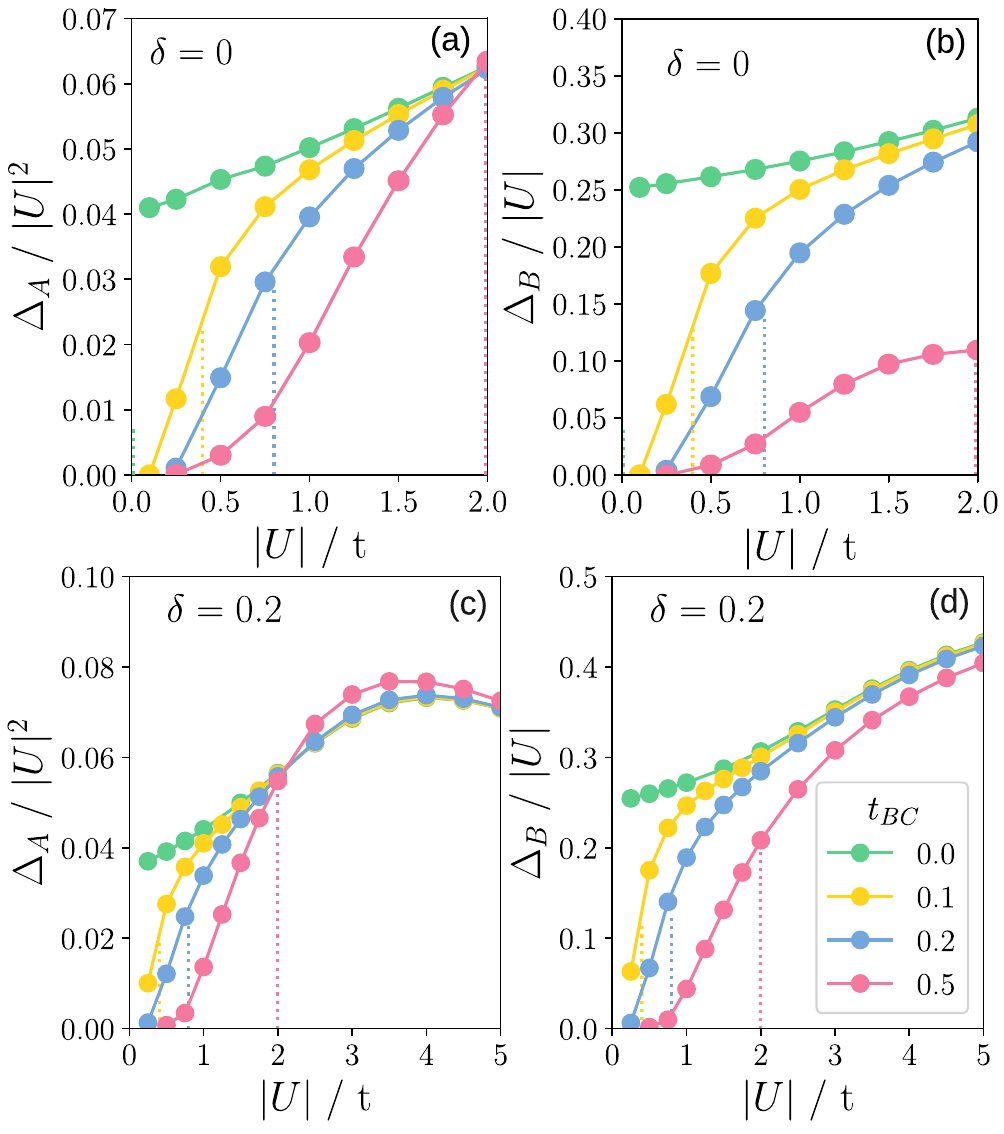} 
\caption{
(a) and (c) (respectively (b) and (d)) show $\Delta_A/|U|^2$ (respectively $\Delta_B/|U|$) as a function of $|U|/t$ for $t_{BC} = 0, 0.1, 0.2$ and $0.5$.
We have considered $\delta=0$ ((a) and (b)) and   $\delta=0.2$ ((c) and (d)).  Due to symmetry $\Delta_C = \Delta_B$. The vertical dotted lines indicate the bandwidth of the QFB (at $|U|=0$).
}
 \label{Fig. 2}
\end{figure} 

\subsection{Pairings}

We now switch on $|U|$. Figure~\ref{Fig. 2} depicts the pairings on orbitals $A$ and $B$ ($C$ being equivalent to $B$) as a function of  $|U|/t$, both in the gapless case ($\delta=0$) and in the gapped case ($\delta=0.2$). 
First, for $t_{BC}=0$, as expected the pairings on $B$ and $C$  are linear with $|U|$ (quadratic on $A$ orbitals) \cite{Peotta_Lieb}. 
As $t_{BC}$ is turned on, the linear behavior at small $|U|$ is rapidly lost and replaced by the usual BCS scaling ($\Delta \sim e^{-1/\rho_F |U|}$) as long as the interaction amplitude is smaller than the bandwidth. For larger $|U|$, pairings increase strongly until ($|U| \gtrsim 10t_{BC}$) they finally coincide with the FB limit.
Recently, it has been reported \cite{Thumin_EPL_2023,Bouzerar_SciPost_2024}  that in half-filled bipartite lattice
and for any non vanishing $|U|$
the pairings obey the following sum-rule $\Sigma = (-\Delta_A+\Delta_B+\Delta_C)/|U|=\frac{1}{2}$.
We find that even if the biparticity is broken when $t_{BC}$ is switched on, the sum-rule still holds with an accuracy larger than $90\%$ when $|U| \gtrsim 10t_{BC}$.
In addition, we also remark that even if we consider relatively large values of $t_{BC}$ (e.g $0.5$), an effective single-band model (not shown) is unable to reproduce, capture the $|U|$ dependence of the pairings plotted in Fig.~\ref{Fig. 2}.
To conclude this section, we find that the pairings are strongly suppressed when $t_{BC}$ is switched on in the weak coupling regime.
\begin{figure}
    \centering
\includegraphics[scale=0.63]{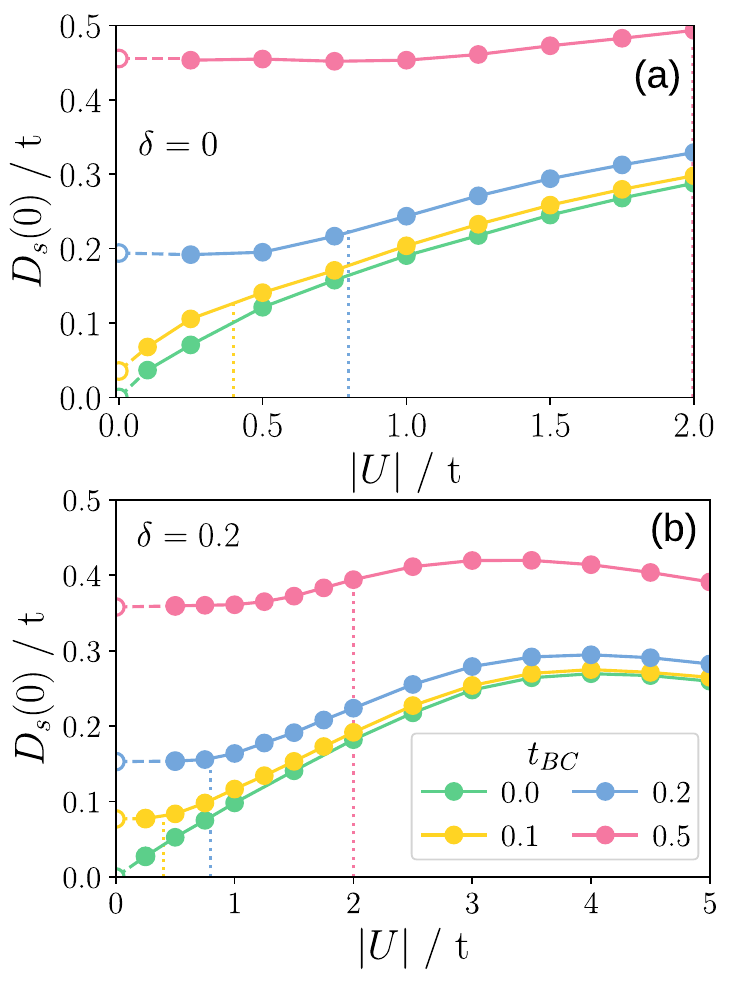}  
     \caption{
     Superfluid weight $D_s$ at $T=0$ as a function of $|U|/t$ for different values of $t_{BC}$. $\delta$ is $0$ (gapless) in the top panel, and $0.2$ (gapped) in the bottom one.
     Vertical dotted lines indicate the value of the bandwidth of the QFB (at $|U|=0$).
     Legend is depicted in panel (b).
     }
 \label{Fig. 3}
\end{figure}


\subsection{Superfluid weight}

In Fig.~\ref{Fig. 3}, the superfluid weight (SFW) is depicted as a function of the interaction strength for different values of $t_{BC}$ with $\delta=0$ (panel (a)) and $\delta=0.2$ (panel (b)). 
First, for $t_{BC}=0$ and in the gapped case, $D_s$ scales linearly with $|U|$ as expected \cite{Peotta_Lieb}. In contrast, for $\delta=0$, $D_s$ is found inconsistent with a linear scaling for small $|U|$. As discussed in Ref.\cite{Torma_revisiting} it scales as $-|U|\ln(|U|)$.
We now consider the case of quasi-FB ($t_{BC} \ne 0$). First, the non-interacting system goes from insulating ($D_s = 0$) to metallic ($D_s\neq0$). As $t_{BC}$ increases the FB acquires a finite bandwidth leading to a finite effective mass $m^* \propto 1/t_{BC}$ which implies $D_s^{U=0} \propto t_{BC}$. 
\\
For any non vanishing $|U|$ and any $\delta$, it is found that the value of $D_s$ in the FB limit is a lower bound. This shows that as $t_{BC}$ grows, the intraband contribution (conventional part) to $D_s$ rapidly grows while the interband-band (geometric) part is unchanged. 
The strong increase in the SFW observed here contrasts with the strong suppression of the pairings depicted in Fig.~\ref{Fig.  2}.
Based on these contradictory results, it is difficult to have a clear intuition on its impact on the critical temperature.

\begin{figure}[h!]
    \centering
\includegraphics[scale=0.57]{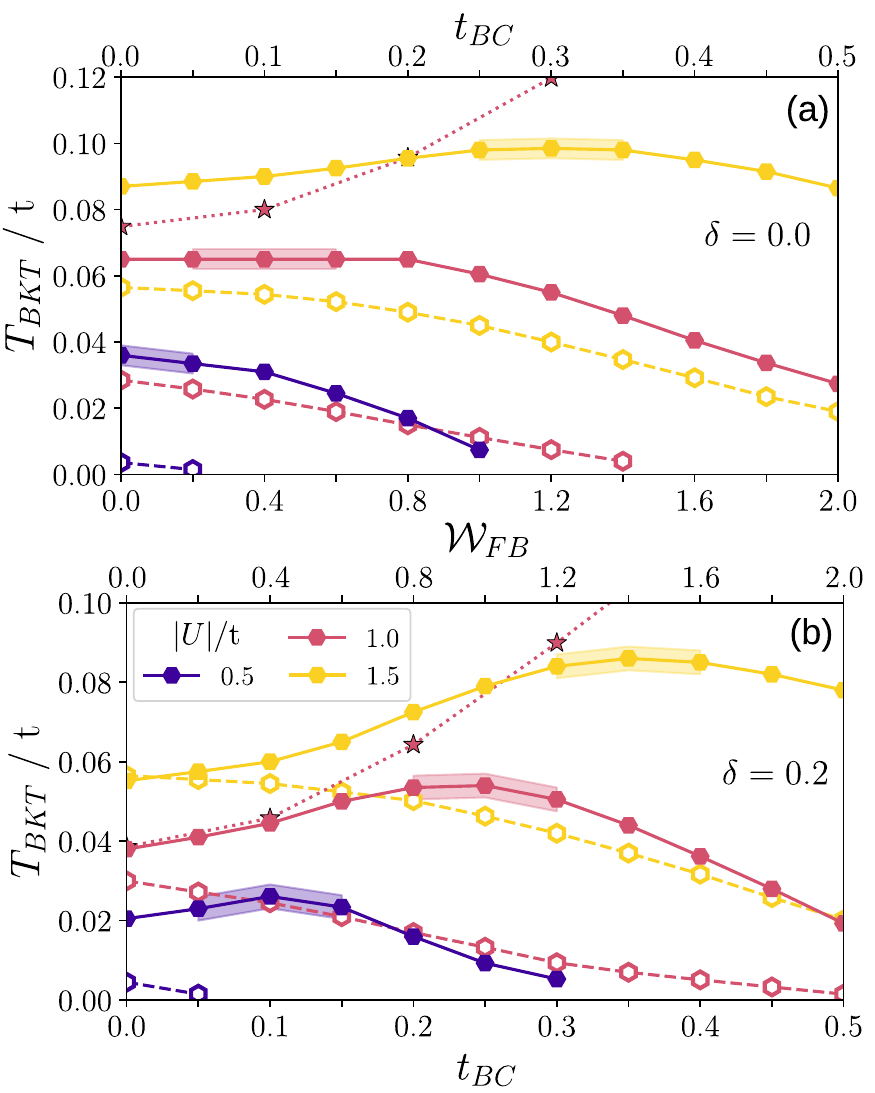}
     \caption{
     $T_{BKT}$ as a function of $t_{BC}$ or bandwidth $\mathcal{W}=4t_{BC}$ for $|U|=0.5, 1.0$ and $1.5$. The upper panel corresponds to $\delta=0$, and bottom one to $\delta=0.2$.
     Dashed (resp. continuous) lines and empty (resp. filled) symbols correspond to $T_{BKT}$ calculated at $\nu=1$ (resp. $\nu=3)$.
     Stars with dotted lines refer to the approximated value of $T_{BKT}$ for $|U|=1$ (see main text). The shaded regions highlight the maximum values of $T_{BKT}$.}
 \label{Fig. 4}
\end{figure}


\subsection{BKT transition temperature}

We now focus on the superconducting transition temperature. In low-dimensional systems ($d<3$), the Mermin-Wagner theorem prevents any phase transition with a spontaneous break of a continuous symmetry at finite temperature \cite{Mermin_Wagner}. However, in two-dimensional materials, a topological phase transition can occur without symmetry breaking yielding a quasi-order at finite temperature known as the Berezinskii–Kosterlitz–Thouless (BKT) transition \cite{Berezinsky_1972, Kosterlitz_1972}. The transition temperature $T_{BKT}$ is defined as follows \cite{Jump_Ds, Peotta_Nature},
\begin{equation}
    D_s(T_{BKT}) = \frac{8}{\pi} k_B T_{BKT}
    .
\end{equation} 
Figure~\ref{Fig. 4} displays the BKT transition temperature as a function of the bandwidth of the QFB ($\mathcal{W}_{FB}=4t_{BC}$) for $|U|/t=0.5, 1.0$ and $1.5$ and at two different carrier density ($\nu=1$ and $3$) in the gapped and gapless case.  
The case $\nu=1$ describes the situation where the lowest dispersive band is half-filled.
For this filling, one observes a suppression of $T_{BKT}$ as $t_{BC}$ increases, in both the gapless and gapped systems.
This reduction is qualitatively well captured considering the BCS expression of $T_{c} \sim e^{-1/\rho_F |U|}$ where the density of states at the Fermi energy $\rho_F$ is found to decay as $t_{BC}$ increases. We remark, that
the observed features are weakly sensitive to $\delta$ since $E_F$ ($\approx -2t$ for small $|U|$) lies far from the region where the dimerization has an impact (vicinity of $E=0$), see Fig.~\ref{Fig. 1}(b,c).
We now set back the filling factor in the QFB at $\nu=3$ , and discuss the gapless and gapped case separately. 
First, for $t_{BC}=0$, the BKT temperature is known to be linear with the interaction strength, allowing higher $T_{BKT}$ than in the case of conventional superconductivity ($\nu=1$).
For $t_{BC}\ne 0$ and $\delta=0$, in contrast to the previous case ($\nu=1$), we now observe in Fig.~\ref{Fig. 4}(a) two different regimes when $|U| \geqslant 1$.
A type of plateau is observed when $\mathcal{W}_{FB} \leqslant |U|$ and even a slight raise for $|U|=1.5$. For larger values of $t_{BC}$, $T_{BKT}$ decreases slowly. On the other, hand for smaller $|U|$, the increase of $t_{BC}$ leads to a rapid suppression of $T_{BKT}$.
The gapped case (Fig.~\ref{Fig. 4}(b)) is even more interesting. As $t_{BC}$ is tuned, the BKT temperature is amplified up to $25\%$ for $|U|=0.5t$ and even $50\%$ for $|U|=1.5t$. The maximum is reached for $\mathcal{W}_{FB} \simeq |U|$. Beyond this value, $T_{BKT}$ decays slowly. 
\\
\\
\begin{figure}[h!]
    \centering
    \includegraphics[scale=0.45]{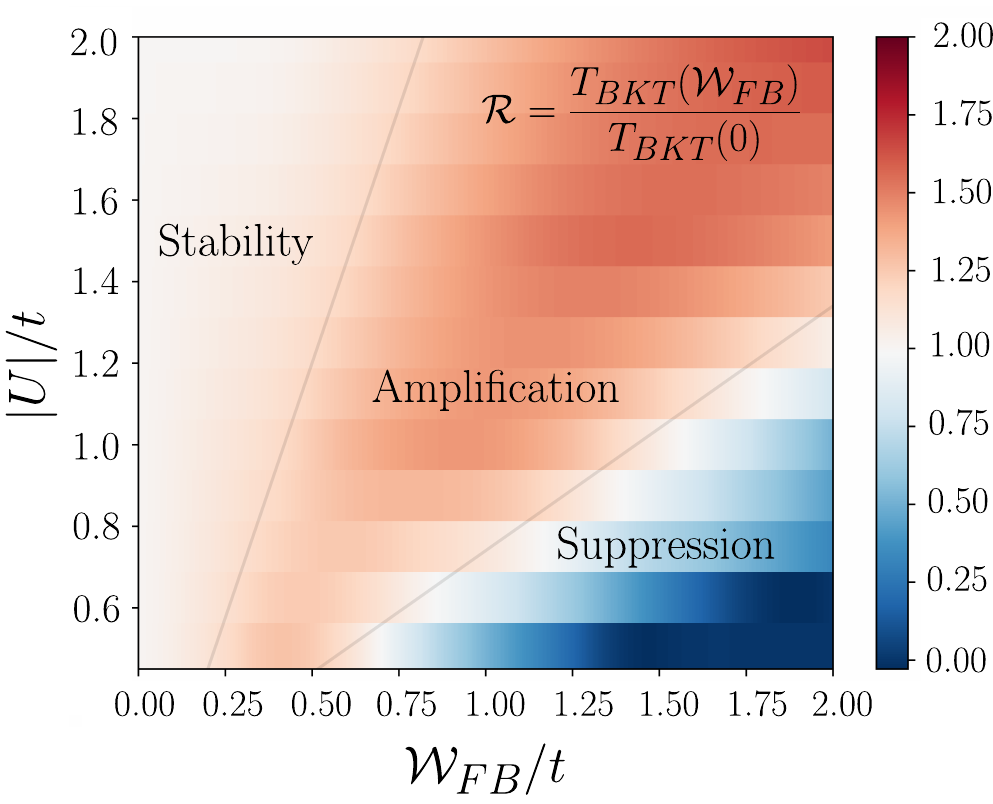} 
    \caption{
    2D color plot for the ratio $\mathcal{R}=T_{BKT}(\mathcal{W}_{FB})/T_{BKT}(0)$ as a function of $\mathcal{W}_{FB}$ and $|U|/t$ for $\delta=0.2$. We clearly observe three different domains: a region of stability ($R\approx 1$), a region of amplification ($\mathcal{R}\gtrsim1$), and a region of suppression $\mathcal{R}\lesssim1$.
    }
    \label{Fig. 5}
\end{figure} 
\\
\\
An important remark must be made in the light of the results of Fig.~\ref{Fig. 3}.
Occasionally, one finds in the literature, estimates of the BKT temperature that use the value of the SFW at $T=0$~K leading to $T_{BKT}^{0}\approx \frac{\pi}{8 k_B} D_s(0)$. This is reasonable in most cases and has the advantage to provide an upper bound.
$T_{BKT}^{0}$ is plotted as a function of the bandwidth of the QFB
in Fig.~\ref{Fig. 4} (dotted stars) for $|U|/t=1$. 
Although a good agreement is found for $t_{BC} \lesssim 0.15t$ in the gapped case, for larger values of the $(B,C)$ hopping, 
$T_{BKT}^{0}$ is both qualitatively and quantitatively incorrect. Indeed, it increases excessively and strongly overestimates $T_{BKT}$. In the gapless case, the situation is even worst since $T_{BKT}^{0}$ fails as well to reproduce the BKT temperature for small values of the bandwidth of the QFB.
\\
To complete and end this section, we show in Fig.~\ref{Fig. 5}, a two dimensional color plot 
of the ratio $\mathcal{R} = T_{BKT}(\mathcal{W}_{FB})/T_{BKT}(0)$, where the variable on the $x$-axis (respectively $y$-axis)
is $\mathcal{W}_{FB}$ (respectively $|U|$). 
As previously discussed, we distinguish three distinct regions. The first corresponds to the cases where $T_{BKT}$ remains unchanged, the second to an increase in the transition temperature by the presence of $t_{BC}$, and finally the third is the one for which $T_{BKT}$ is suppressed. 
These results are to be compared with those of Fig.~\ref{Fig. 2}. One could have concluded that $t_{BC}$ would have a dramatic impact on the critical temperatures because it leads to the drastic suppression of the pairings. However, one clearly observe that the $T_{BKT}$ can be strongly reinforced by its presence.

\subsection{Coherence length}

\begin{figure}[h!]
    \centering
\includegraphics[scale=0.57]{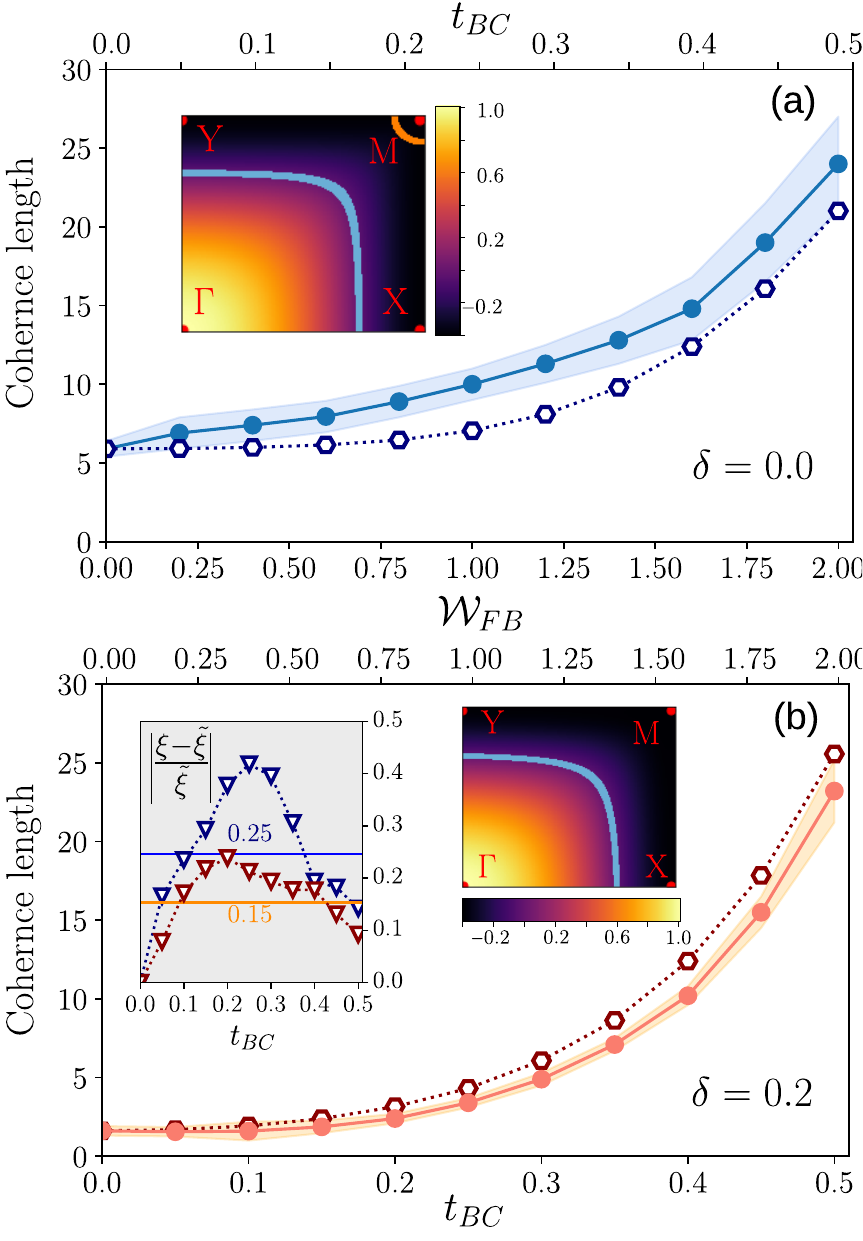}
     \caption{
     Coherence length $\xi$ (filled symbols) as a function of $t_{BC}$ for $|U|/t=1$ and $\delta=0$ in (a) and $\delta=0.2$ in (b). Shaded areas correspond to error bars. Open symbols represent the semi-analytic formula $\Tilde{\xi}$ (see text). In both cases, the inset depicts the QFB and the Fermi Surface at $\nu=3$ in the upper quarter BZ.
     The second inset in (b) represents the deviation between $\xi$ and $\Tilde{\xi}$ when $\delta=0$ (blue triangles) and $\delta=0.2$ (red triangles), and the average values of the deviation (continuous lines).
     }
 \label{Fig. 6}
\end{figure}

Let us now discuss the coherence length in quasi-flat band systems, a topic often addressed in recent studies \cite{Iskin_coherence_length,Law_GL,Thumin_coherence_length,Law_anomalous}. The coherence length is defined as the  length scale that characterizes the exponential decay of the anomalous correlation function,
$K_\lambda(\textbf{r})=\langle \hat{c}_{i\lambda}^\dagger \hat{c}_{j\lambda}^\dagger \rangle \sim e^{-|\textbf{r}|/\xi_\lambda}$, where $\textbf{r}=\textbf{r}_i-\textbf{r}_j$ \cite{Thumin_coherence_length}. Note that  $\xi_\lambda$ was numerically found $\lambda$-independent, hence the orbital index is dropped.
Figure \ref{Fig. 6} displays the coherence length in the half-filled system as a function of $t_{BC}$ for $|U|/t = 1$, $\delta = 0$ and $0.2$.
First, we observe for both $\delta=0$ and $\delta=0.2$ the same qualitative behavior (monotonic increase), which could be understood by the increase of Fermi velocity. This will be clarified in what follows. More precisely, one observes two different regimes. For $t_{BC} \le 0.2$, $\xi$ is found to vary weakly with $t_{BC}$ and beyond, it grows rapidly.
In addition, at $t_{BC} = 0$, we find $\xi=5.9a$ for $\delta=0$ and $\xi=1.62a$ for $\delta=0.2$. The presence of the gap leads to a strong reduction of $\xi$,
a trend similar to what was observed for the SFW \cite{Torma_revisiting, Batrouni_Designer_Flat_Bands}.  
These values are in contrast to those found in the case of conventional superconductivity, for which very large values of $\xi$ are generally expected (up to $10^2-10^3$a).
\\
Coherence length in FB superconductors is a very controversial topic that has often been discussed recently. In recent analytical studies,
it has been suggested that the coherence length in quasi FB has the form,
\begin{eqnarray}
  \xi=\sqrt{\langle g \rangle + \xi_{BCS}^2},
  \label{xi-eq}
\end{eqnarray}
the first term being the quantum metric (geometric contribution) and the second one the conventional part \cite{Iskin_coherence_length,Law_GL,Thumin_coherence_length,Law_anomalous}.
The quantum metric (QM) reads $\langle g \rangle = \int \frac{d^2k}{(2\pi)^2} \langle \partial_x \psi^{FB}_{\textbf{k}} | (1- |  \psi^{FB}_{\textbf{k}} \rangle \langle \psi^{FB}_{\textbf{k}}| ) |\partial_x \psi^{FB}_{\textbf{k}} \rangle$, $\partial_x$ stands for $\partial/\partial k_x$, and $|\psi^{FB}_{\textbf{k}} \rangle$ is the FB eigenstate at $U=0$. $\xi_{BCS}=\hbar v_F/ \Delta$ being the coherence length of a conventional superconductor with gap $\Delta$ and Fermi velocity $v_F$. 
In the case where the FB is exactly flat, $v_F$ vanishes and hence one expects $\xi = \sqrt{\langle g \rangle}$ .
However, in a recent detailed numerical study, where various flat band systems were considered, it has been found that this expression does not hold \cite{Thumin_coherence_length}. 
\\
In order to understand how a finite value of $t_{BC}$ affects the coherence length, we propose to compute a semi-analytic formula inspired from Eq.~\eqref{xi-eq} where we replace $\sqrt{g}$ by the numerical value of the coherence length at $t_{BC}=0$, $\xi_{FB}=\xi(t_{bc}=0)$,
\begin{equation}
    \Tilde{\xi} = \sqrt{\xi_{FB}^2 + \xi_{BCS}^2} 
    .
    \label{xi_sa}
\end{equation}
In $\xi_ {BCS}$, $\Delta$ is replaced by $\Delta_{avg}=\sum_\lambda \Delta_\lambda/3$, and $v_F^2 =  \langle \textbf{v}_{k}^2 \rangle_{FS}$ is averaged over the Fermi surface of the QFB at $|U|=0$.
$\Tilde{\xi}$ is depicted with the empty symbols in Fig.~\ref{Fig. 6} (a) and (b). Clearly, the full numerical calculation and the semi-analytic one agree very well. More quantitatively, it is found that the relative deviation between $\xi$ and $\Tilde{\xi}$ (depicted in inset of the panel (b)) is about
$15\%$ only in average for the gapped case and $25\%$ for the gapless system. The better agreement obtained in gapped case can be explained by the nature of the Fermi surface (FS) of the system shown in inset of Fig.~\ref{Fig. 6} (a) and (b).
While the QFB dispersion is similar for both $\delta=0$ and $0.2$, in the gapped case, the FS possesses only a single contribution originating from the QFB. In the gapless case there is also electron pockets centered around the M-point of the Brillouin zone coming from the upper band.
This could be anticipated
from the dispersions shown in Fig.~\ref{Fig. 1}(b).


\subsection{Breaking the biparticity with $(A,A)$ hopping}
\begin{figure}[h!]
    \centering
    \includegraphics[scale=0.55]{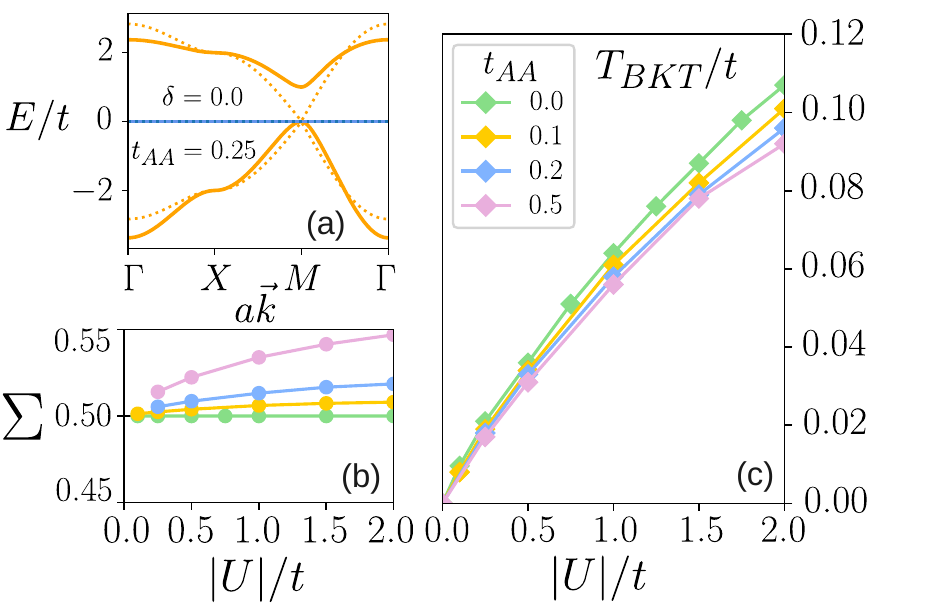} 
    \caption{
    (a) Impact of $(A,A)$ hopping on the dispersions (continuous lines), when $t_{AA}=0.25~t$.
    (b) (respectively (c)) $\Sigma=\frac{1}{|U|}(-\Delta_{A} + \Delta_{B} +\Delta_{C})$ (respectively $T_{BKT}$) as a function of $|U|$ in the half-filled lattice for different values of $t_{AA}$. The dimerization parameter is $\delta = 0$.
    }
    \label{Fig. 7}
\end{figure}

Before turning to the last section of this work, we briefly analyze the effect of a $(A,A)$ hopping. Up to this point, the biparticity was broken by a $(B,C)$ hopping yielding a QFB. One can naturally ask about the effect of a non-zero $t_{AA}$. The non-interacting perturbed Hamiltonian reads,
\begin{equation}
    h_\textbf{k}^\sigma = 
    \begin{bmatrix}
        f_{AA} &  f_x^* & f_y^* \\ 
        f_x    &      0 & 0     \\
        f_y    &      0 & 0     \\
    \end{bmatrix}
    ,
\end{equation}
with $f_{AA}=-2t_{AA}(\cos{k_xa}+\cos{k_ya})$.
First, one can straightforwardly see that the CLS is not affected by a finite $t_{AA}$, thus the FB Bloch eigenstate and therefore the QM remain unchanged. 
The dispersion for $t_{AA}=0.25$
is depicted in Fig.~\ref{Fig. 7} (a). We see that the lower dispersive band and the FB still touch at point M, but now in a quadratic manner. The upper dispersive band is found significantly narrower, a gap opens up separating it from the FB.
Numerically, for a given value of $|U|$, we have found that $t_{AA}$ has only a minor impact on $D_s(0)$. The effect is even smaller on $T_{BKT}$ plotted in Fig.~\ref{Fig. 7}(c).
In contrast to a $t_{BC}$ perturbation, when $t_{AA}$ is switched on, no increase of $T_{BKT}$ is found but the reduction of $T_{BKT}$ stays very limited. We briefly discuss how
the sum-rule $\Sigma = (-\Delta_A+\Delta_B+\Delta_C)/|U|=
\frac{1}{2}$ \cite{Thumin_EPL_2023,Bouzerar_SciPost_2024} is modified in the presence of $t_{AA}$ which as well breaks the bipartite character without affecting the FB eigenstates. In Fig.~\ref{Fig. 7}(c) $\Sigma$ is plotted as a function of $|U|$, for different values of $t_{AA}$.
As can be seen, the sum-rule appears relatively robust (minor impact), even when $t_{AA}$ reaches large values ($0.5$). In addition, we remark as well that $\Sigma$ is slightly larger than $1/2$.
To conclude this section, one can understand the small impact of $t_{AA}$ because (i) the FB remains strictly flat and (ii) the associated eigenstates are unaffected. 

\subsection{Effect of the reduction of dimensionality: the sawtooth chain}

\begin{figure}[h!]
    \centering
    \includegraphics[scale=0.5]{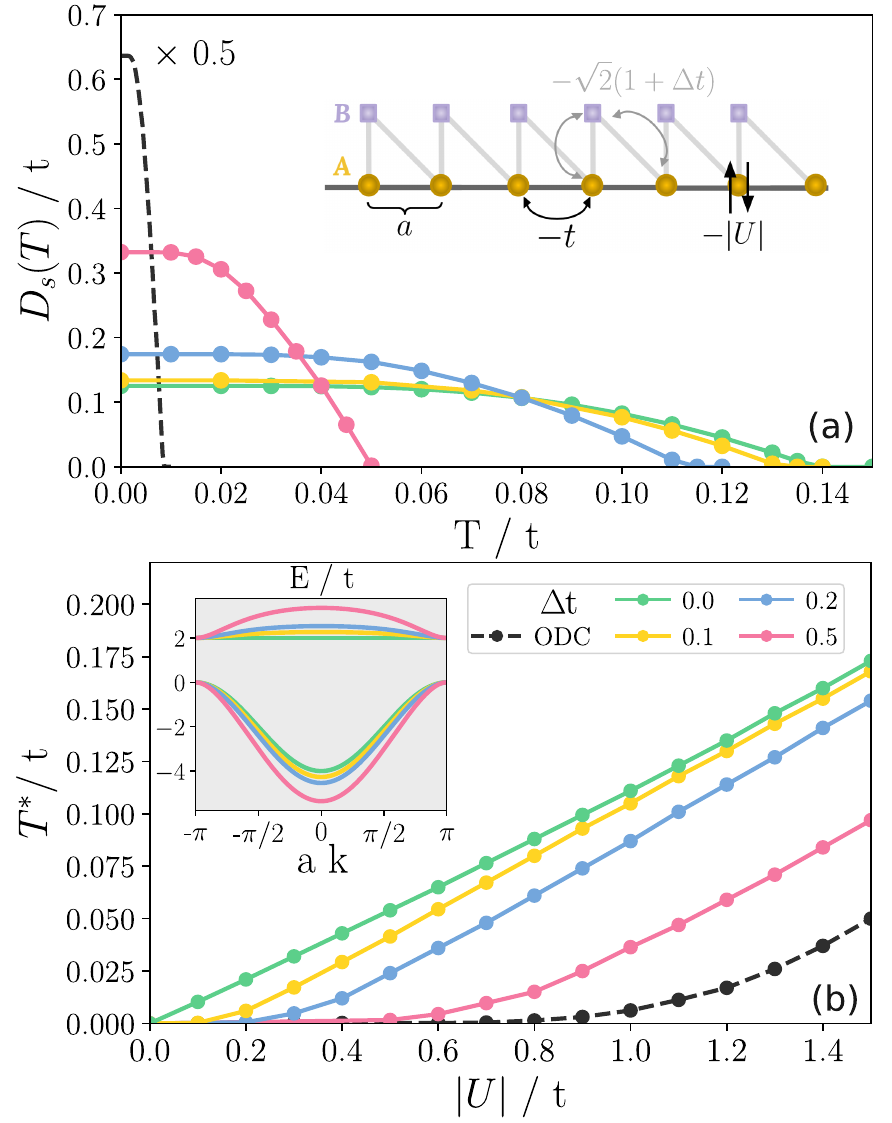} 
    \caption{
    \textbf{(a)} Superfluid weight $D_s$ as a function of temperature in the half-filled sawtooth chain as illustrated in the inset, for $|U|/t = 1$ and $\Delta t=0, 0.1, 0.2$ and $0.5$. ODC refers to the half-filled standard one-dimensional chain corresponding to $\Delta t=-1$.  In the ODC case, $D_s$ is multiplied by $0.5$.
    \textbf{(b)} Characteristic temperature $T^*$ (see text) as a function of $|U|/t$ for the same values of $\Delta t$. 
    The dispersions in the non-interacting case are depicted in the inset. 
    }
    \label{Fig. 8}
\end{figure}

The purpose of this last section is to verify whether the robustness of the FB superconductivity when spatial dimensionality of the system is reduced. For this we propose to study the case of the sawtooth chain depicted in inset of Fig.~\ref{Fig. 8}(a). In the basis $\{\hat{c}_{Ak\sigma}^{\dagger},\hat{c}_{Bk\sigma}^{\dagger}\}$, the single particle Hamiltonian reads, 
\begin{equation}
    h_k^\sigma = 
    \begin{bmatrix}
        f_{AA} & f_{BC}^* \\ 
        f_{BC} &        0 \\
    \end{bmatrix}
    ,
\end{equation}
where $f_{AA} = -2t\cos{ka}$, and $f_{AB}=-\sqrt{2}(1+\Delta t)(1+e^{ika})$.
The spectrum (depicted in inset of Fig.~\ref{Fig. 8}(b)) host a FB at $E=2t$ when the $(A,B)$ hopping is $t_{AB}=-\sqrt{2}t$. The FB and the dispersive bands are separated by a large gap $2t$. 
To turn the FB weakly dispersive, we add a small perturbation $\Delta t$ to the $(A,B)$ hopping, $t_{AB}=-\sqrt{2}(t+\Delta t)$. 
Note that in contrast to the Lieb lattice the sawtooth chain is not bipartite and the CLS has a finite weight on both $A$ and $B$ sublattice. In what follows, the filling factor is set to $\nu=3$ (half-filled QFB).
Figure~\ref{Fig. 8}(a) displays the SFW as a function of temperature for $\Delta t$ ranging from $0$ (FB) to $0.5$. We have as well included the case $\Delta t=-1$ which corresponds to the standard one dimensional chain (ODC).
At $T=0$, as $\Delta t$ increases the SFW increases rapidly due to the growing of the intraband contribution.
On the other hand, the mean-field critical temperature is dramatically suppressed. In particular, the critical temperature of the ODC is $0.0085~t$ compared to $0.14~t$ for $\Delta t=0$.
The system being one-dimensional no phase transition can occur at finite temperature, however it is instructive to define a characteristic temperature for the thermal fluctuations $T^*$ defined as $D_s(T^*)=D_s(0)/2$ \cite{Thumin_PRB_2023}. 
In Fig~\ref{Fig. 8}(b), $T^*$ is depicted as a function of $|U|$ for the same values of $\Delta t$. 
First, we immediately observe that for any value of $|U|$, $T^*$ decreases monotonously as $\Delta t$ grows and has an upper bound corresponding to the flat band limit.
For $|U|$ larger than the bandwidth, we observe that $T^*$ scales linearly with the interaction strength. In addition as $\Delta t$ increases for a fixed $U$, $T^*$ is strongly suppressed when $|\Delta t| \gtrsim |U|/2$.
\\
To conclude this section, the FB superconducting phase in one-dimensional systems clearly seems less stable in the face of perturbations. This is understandable, as one-dimensional systems are much more sensitive to quantum and thermal fluctuations than higher-dimensional systems.
In particular, in contrast to the gapped Lieb lattice, we do not observe here any possibility of reinforcement of the superconducting phase.
It might be interesting to perform similar calculations in a one-dimensional bipartite lattice, such as the stub lattice, to check whether the bipartite character (before perturbation) protects the system against thermal fluctuations.

\section{Conclusion}
In conclusion, we have studied the stability of the superconducting phase in flat bands against perturbations  in both the Lieb lattice and the sawtooth chain.
It has been found that $T_{BKT}$ is robust against perturbations and can even be significantly reinforced in the case of the Lieb lattice. 
It is also shown that the calculated coherence length obeys a modified version of a recently proposed equation that contains both a geometric contribution and an intraband part.
On the other hand, in reduced dimensionality, due to the importance of quantum and thermal fluctuations, it appears that the superconducting phase in flat bands is much less stable against perturbations.
From a practical point of view, our results could be of great interest for the experimental exploration of superconductivity in two-dimensional flat band systems, as they suggest that low dispersion of quasi-FBs is not  detrimental for observing superconductivity of a geometric nature.

\vspace{1cm}


\begin{thebibliography}{10}

\bibitem{BCS}
J.~Bardeen, L.~N. Cooper, and J.~R. Schrieffer, ``Theory of superconductivity,'' {\em Phys. Rev.}, vol.~108, pp.~1175--1204, Dec 1957.

\bibitem{Dupuis}
N.~Dupuis, ``Berezinskii-kosterlitz-thouless transition and bcs-bose crossover in the two-dimensional attractive hubbard model,'' {\em Phys. Rev. B}, vol.~70, p.~134502, Oct 2004.

\bibitem{Gap_linear_1990}
V.~A. {Khodel'} and V.~R. {Shaginyan}, ``{Superfluidity in system with fermion condensate},'' {\em ZhETF Pisma Redaktsiiu}, vol.~51, p.~488, May 1990.

\bibitem{Gap_linear_1994}
V.~Khodel, V.~Shaginyan, and V.~Khodel, ``New approach in the microscopic fermi systems theory,'' {\em Physics Reports}, vol.~249, no.~1, pp.~1--134, 1994.

\bibitem{Batrouni_CuO2}
V.~I. Iglovikov, F.~H\'ebert, B.~Gr\'emaud, G.~G. Batrouni, and R.~T. Scalettar, ``Superconducting transitions in flat-band systems,'' {\em Phys. Rev. B}, vol.~90, p.~094506, Sep 2014.

\bibitem{Aoki2020}
H.~Aoki, ``Theoretical possibilities for flat band superconductivity,'' {\em Journal of Superconductivity and Novel Magnetism}, vol.~33, pp.~2341--2346, Aug 2020.

\bibitem{BCS_FB}
S.~Miyahara, S.~Kusuta, and N.~Furukawa, ``Bcs theory on a flat band lattice,'' {\em Physica C: Superconductivity}, vol.~460-462, pp.~1145--1146, 2007.
\newblock Proceedings of the 8th International Conference on Materials and Mechanisms of Superconductivity and High Temperature Superconductors.

\bibitem{Batrouni_sawtooth}
S.~M. Chan, B.~Gr\'emaud, and G.~G. Batrouni, ``Pairing and superconductivity in quasi-one-dimensional flat-band systems: Creutz and sawtooth lattices,'' {\em Phys. Rev. B}, vol.~105, p.~024502, Jan 2022.

\bibitem{Volovik_T_linear}
N.~B. Kopnin, T.~T. Heikkil\"a, and G.~E. Volovik, ``High-temperature surface superconductivity in topological flat-band systems,'' {\em Phys. Rev. B}, vol.~83, p.~220503, Jun 2011.

\bibitem{Peotta_Nature}
S.~Peotta and P.~T{\"o}rm{\"a}, ``Superfluidity in topologically nontrivial flat bands,'' {\em Nature Communications}, vol.~6, p.~8944, Nov 2015.

\bibitem{Peotta_Lieb}
A.~Julku, S.~Peotta, T.~I. Vanhala, D.-H. Kim, and P.~T\"orm\"a, ``Geometric origin of superfluidity in the lieb-lattice flat band,'' {\em Phys. Rev. Lett.}, vol.~117, p.~045303, Jul 2016.

\bibitem{Iskin}
M.~Iskin, ``Cooper pairing, flat-band superconductivity, and quantum geometry in the pyrochlore-hubbard model,'' {\em Phys. Rev. B}, vol.~109, p.~174508, May 2024.

\bibitem{Provost_metric}
J.~P. Provost and G.~Vallee, ``Riemannian structure on manifolds of quantum states,'' {\em Communications in Mathematical Physics}, vol.~76, pp.~289--301, Sep 1980.

\bibitem{Berry_5_years}
M.~V. Berry, ``{THE QUANTUM PHASE, FIVE YEARS AFTER},'' 1989.

\bibitem{Thumin_boost_QM}
M.~Thumin and G.~Bouzerar, ``Strengthening of the superconductivity by real-space decimation of the flat-band states,'' {\em Phys. Rev. B}, vol.~110, p.~134512, Oct 2024.

\bibitem{Kagome_optical_lattice}
G.-B. Jo, J.~Guzman, C.~K. Thomas, P.~Hosur, A.~Vishwanath, and D.~M. Stamper-Kurn, ``Ultracold atoms in a tunable optical kagome lattice,'' {\em Phys. Rev. Lett.}, vol.~108, p.~045305, Jan 2012.

\bibitem{Lieb_optical_lattice}
S.~Taie, H.~Ozawa, T.~Ichinose, T.~Nishio, S.~Nakajima, and Y.~Takahashi, ``Coherent driving and freezing of bosonic matter wave in an optical lieb lattice,'' {\em Science Advances}, vol.~1, no.~10, p.~e1500854, 2015.

\bibitem{Lieb_optical_lattice2}
H.~Ozawa, S.~Taie, T.~Ichinose, and Y.~Takahashi, ``Interaction-driven shift and distortion of a flat band in an optical lieb lattice,'' {\em Phys. Rev. Lett.}, vol.~118, p.~175301, Apr 2017.

\bibitem{Sawtooth_optical_lattice}
F.~A. An, E.~J. Meier, and B.~Gadway, ``Engineering a flux-dependent mobility edge in disordered zigzag chains,'' {\em Phys. Rev. X}, vol.~8, p.~031045, Aug 2018.

\bibitem{photonic_Lieb_Kagome_Honeycomb}
L.~Tang, D.~Song, S.~Xia, S.~Xia, J.~Ma, W.~Yan, Y.~Hu, J.~Xu, D.~Leykam, and Z.~Chen, ``Photonic flat-band lattices and unconventional light localization,'' {\em Nanophotonics}, vol.~9, no.~5, pp.~1161--1176, 2020.

\bibitem{photonic_metallic_Kagome}
Y.~Nakata, T.~Okada, T.~Nakanishi, and M.~Kitano, ``Observation of flat band for terahertz spoof plasmons in a metallic kagom\'e lattice,'' {\em Phys. Rev. B}, vol.~85, p.~205128, May 2012.

\bibitem{photonic_metallic_Lieb}
S.~Kajiwara, Y.~Urade, Y.~Nakata, T.~Nakanishi, and M.~Kitano, ``Observation of a nonradiative flat band for spoof surface plasmons in a metallic lieb lattice,'' {\em Phys. Rev. B}, vol.~93, p.~075126, Feb 2016.

\bibitem{TaS2_exp}
Y.~D. Wang, W.~L. Yao, Z.~M. Xin, T.~T. Han, Z.~G. Wang, L.~Chen, C.~Cai, Y.~Li, and Y.~Zhang, ``Band insulator to mott insulator transition in 1t-tas2,'' {\em Nature Communications}, vol.~11, p.~4215, Aug 2020.

\bibitem{TaS2_th}
H.~Bae, R.~Valenti, I.~I. Mazin, and B.~Yan, ``Designing flat bands, localized and itinerant states in tas2 trilayer heterostructures,'' 2025.

\bibitem{FB_WSe2}
L.~Wang, E.-M. Shih, A.~Ghiotto, L.~Xian, D.~A. Rhodes, C.~Tan, M.~Claassen, D.~M. Kennes, Y.~Bai, B.~Kim, K.~Watanabe, T.~Taniguchi, X.~Zhu, J.~Hone, A.~Rubio, A.~N. Pasupathy, and C.~R. Dean, ``Correlated electronic phases in twisted bilayer transition metal dichalcogenides,'' {\em Nature Materials}, vol.~19, pp.~861--866, Aug 2020.

\bibitem{MoS2_WS2_WSe2}
M.~H. Naik and M.~Jain, ``Ultraflatbands and shear solitons in moir\'e patterns of twisted bilayer transition metal dichalcogenides,'' {\em Phys. Rev. Lett.}, vol.~121, p.~266401, Dec 2018.

\bibitem{heterobilayers}
D.~A. Ruiz-Tijerina and V.~I. Fal'ko, ``Interlayer hybridization and moir\'e superlattice minibands for electrons and excitons in heterobilayers of transition-metal dichalcogenides,'' {\em Phys. Rev. B}, vol.~99, p.~125424, Mar 2019.

\bibitem{III-V_SC}
N.~A. Franchina~Vergel, L.~C. Post, D.~Sciacca, M.~Berthe, F.~Vaurette, Y.~Lambert, D.~Yarekha, D.~Troadec, C.~Coinon, G.~Fleury, G.~Patriarche, T.~Xu, L.~Desplanque, X.~Wallart, D.~Vanmaekelbergh, C.~Delerue, and B.~Grandidier, ``Engineering a robust flat band in iii–v semiconductor heterostructures,'' {\em Nano Letters}, vol.~21, no.~1, pp.~680--685, 2021.
\newblock PMID: 33337891.

\bibitem{Cao2018_1}
Y.~Cao, V.~Fatemi, S.~Fang, K.~Watanabe, T.~Taniguchi, E.~Kaxiras, and P.~Jarillo-Herrero, ``Unconventional superconductivity in magic-angle graphene superlattices,'' {\em Nature}, vol.~556, pp.~43--50, Apr 2018.

\bibitem{Cao2018_2}
Y.~Cao, V.~Fatemi, A.~Demir, S.~Fang, S.~L. Tomarken, J.~Y. Luo, J.~D. Sanchez-Yamagishi, K.~Watanabe, T.~Taniguchi, E.~Kaxiras, R.~C. Ashoori, and P.~Jarillo-Herrero, ``Correlated insulator behaviour at half-filling in magic-angle graphene superlattices,'' {\em Nature}, vol.~556, pp.~80--84, Apr 2018.

\bibitem{FB_supra_WSe2}
Y.~Guo, J.~Pack, J.~Swann, L.~Holtzman, M.~Cothrine, K.~Watanabe, T.~Taniguchi, D.~G. Mandrus, K.~Barmak, J.~Hone, A.~J. Millis, A.~Pasupathy, and C.~R. Dean, ``Superconductivity in 5.0{\textdegree} twisted bilayer wse2,'' {\em Nature}, vol.~637, pp.~839--845, Jan 2025.

\bibitem{proximity_effect_FB}
A.~Diez-Carlon, J.~Diez-Merida, P.~Rout, D.~Sedov, P.~Virtanen, S.~Banerjee, R.~P.~S. Penttila, P.~Altpeter, K.~Watanabe, T.~Taniguchi, S.~Y. Yang, K.~T. Law, T.~T. Heikkila, P.~Torma, M.~S. Scheurer, and D.~K. Efetov, ``Probing the flat-band limit of the superconducting proximity effect in twisted bilayer graphene josephson junctions,'' 2025.

\bibitem{Lieb_Lattice_Origine}
E.~H. Lieb, ``Two theorems on the hubbard model,'' {\em Phys. Rev. Lett.}, vol.~62, pp.~1201--1204, Mar 1989.

\bibitem{Bound_inequality_roy}
R.~Roy, ``Band geometry of fractional topological insulators,'' {\em Phys. Rev. B}, vol.~90, p.~165139, Oct 2014.

\bibitem{Hofmann_QFB}
J.~S. Hofmann, E.~Berg, and D.~Chowdhury, ``Superconductivity, pseudogap, and phase separation in topological flat bands,'' {\em Phys. Rev. B}, vol.~102, p.~201112, Nov 2020.

\bibitem{Torma_band_geometry_berry}
L.~Liang, T.~I. Vanhala, S.~Peotta, T.~Siro, A.~Harju, and P.~T\"orm\"a, ``Band geometry, berry curvature, and superfluid weight,'' {\em Phys. Rev. B}, vol.~95, p.~024515, Jan 2017.

\bibitem{Batrouni_Designer_Flat_Bands}
S.~M. Chan, B.~Gr\'emaud, and G.~G. Batrouni, ``Designer flat bands: Topology and enhancement of superconductivity,'' {\em Phys. Rev. B}, vol.~106, p.~104514, Sep 2022.

\bibitem{Hofmann_PRB}
J.~S. Hofmann, E.~Berg, and D.~Chowdhury, ``Superconductivity, charge density wave, and supersolidity in flat bands with a tunable quantum metric,'' {\em Phys. Rev. Lett.}, vol.~130, p.~226001, May 2023.

\bibitem{Chi_QMC}
J.~S. Hofmann, E.~Berg, and D.~Chowdhury, ``Superconductivity, charge density wave, and supersolidity in flat bands with a tunable quantum metric,'' {\em Phys. Rev. Lett.}, vol.~130, p.~226001, May 2023.

\bibitem{Peri_PRL}
V.~Peri, Z.-D. Song, B.~A. Bernevig, and S.~D. Huber, ``Fragile topology and flat-band superconductivity in the strong-coupling regime,'' {\em Phys. Rev. Lett.}, vol.~126, p.~027002, Jan 2021.

\bibitem{Arbeitman_PRL}
J.~Herzog-Arbeitman, V.~Peri, F.~Schindler, S.~D. Huber, and B.~A. Bernevig, ``Superfluid weight bounds from symmetry and quantum geometry in flat bands,'' {\em Phys. Rev. Lett.}, vol.~128, p.~087002, Feb 2022.

\bibitem{Thumin_coherence_length}
M.~Thumin and G.~Bouzerar, ``Correlation functions and characteristic lengthscales in flat band superconductors,'' {\em arXiv:2405.06215}.

\bibitem{Torma_revisiting}
K.-E. Huhtinen, J.~Herzog-Arbeitman, A.~Chew, B.~A. Bernevig, and P.~T\"orm\"a, ``Revisiting flat band superconductivity: Dependence on minimal quantum metric and band touchings,'' {\em Phys. Rev. B}, vol.~106, p.~014518, Jul 2022.

\bibitem{Th_Lieb_uniform}
E.~H. Lieb, M.~Loss, and R.~J. McCann, ``Uniform density theorem for the hubbard model,'' {\em Journal of Mathematical Physics}, vol.~34, no.~3, pp.~891--898, 1993.

\bibitem{Thumin_EPL_2023}
M.~Thumin and G.~Bouzerar, ``Constraint relations for superfluid weight and pairings in a chiral flat band superconductor,'' {\em Europhysics Letters}, vol.~144, p.~56001, dec 2023.

\bibitem{Bouzerar_SciPost_2024}
G.~Bouzerar and M.~Thumin, ``{Hidden symmetry of Bogoliubov de Gennes quasi-particle eigenstates and universal relations in flat band superconducting bipartite lattices},'' {\em SciPost Phys. Core}, vol.~7, p.~018, 2024.

\bibitem{Mermin_Wagner}
N.~D. Mermin and H.~Wagner, ``Absence of ferromagnetism or antiferromagnetism in one- or two-dimensional isotropic heisenberg models,'' {\em Phys. Rev. Lett.}, vol.~17, pp.~1133--1136, Nov 1966.

\bibitem{Berezinsky_1972}
V.~L. Berezinsky, ``{Destruction of Long-range Order in One-dimensional and Two-dimensional Systems Possessing a Continuous Symmetry Group. II. Quantum Systems.},'' {\em Sov. Phys. JETP}, vol.~34, no.~3, p.~610, 1972.

\bibitem{Kosterlitz_1972}
J.~M. Kosterlitz and D.~J. Thouless, ``Long range order and metastability in two dimensional solids and superfluids. (application of dislocation theory),'' {\em Journal of Physics C: Solid State Physics}, vol.~5, p.~L124, jun 1972.

\bibitem{Jump_Ds}
D.~R. Nelson and J.~M. Kosterlitz, ``Universal jump in the superfluid density of two-dimensional superfluids,'' {\em Phys. Rev. Lett.}, vol.~39, pp.~1201--1205, Nov 1977.

\bibitem{Iskin_coherence_length}
M.~Iskin, ``Coherence length and quantum geometry in a dilute flat-band superconductor,'' {\em Phys. Rev. B}, vol.~110, p.~144505, Oct 2024.

\bibitem{Law_GL}
S.~A. Chen and K.~T. Law, ``Ginzburg-landau theory of flat-band superconductors with quantum metric,'' {\em Phys. Rev. Lett.}, vol.~132, p.~026002, Jan 2024.

\bibitem{Law_anomalous}
J.-X. Hu, S.~A. Chen, and K.~T. Law, ``Anomalous coherence length in superconductors with quantum metric,'' {\em Communications Physics}, vol.~8, p.~20, Jan 2025.

\bibitem{Thumin_PRB_2023}
M.~Thumin and G.~Bouzerar, ``Flat-band superconductivity in a system with a tunable quantum metric: The stub lattice,'' {\em Phys. Rev. B}, vol.~107, p.~214508, Jun 2023.

\end{thebibliography}

\end{document}